\documentclass{article}
\usepackage[utf8]{inputenc}
\usepackage{textcomp}
\usepackage{graphicx} 
\usepackage{authblk} 
\usepackage{geometry}
\usepackage{float}
\usepackage{amsmath}
\usepackage{amsfonts}
\usepackage{multirow}
\usepackage[colorlinks=true, allcolors=blue]{hyperref}
\usepackage{orcidlink}
\usepackage{float}
\usepackage{tabularx}  
\usepackage{booktabs}  
\usepackage{cite}

\usepackage{xcolor}

\title{Statistical Constraints on Anisotropic Bianchi-III Cosmology in $f(R,T)$-Gravity Using MCMC Methods}

\author[1]{Mayur Mune\orcidlink{0009-0009-2170-8909}}
\author[2]{Praveen Kumar Dhankar\orcidlink{0000-0002-8201-6019}}
\author[3]{Safiqul Islam\orcidlink{0000-0003-1373-4137} }
\author[4]{Behnam Pourhassan\orcidlink{0000-0003-1338-7083}}
\author[5]{Muhammad Aamir\orcidlink{0000-0002-5782-0064}}
\author[6]{Faisal Haroon\orcidlink{0000-0003-4353-6318}}

\affil[1,2]{{\scriptsize Symbiosis Institute of Technology, Nagpur Campus, Symbiosis International (Deemed University), Pune 440008, Maharastra, India. $^1$Email: mayurmune1999@gmail.com; $^2$Email: pkumar6743@gmail.com }}

\affil[3]{{{\scriptsize Department of Basic Sciences, General Administration of Preparatory Year, King Faisal University, P.O. Box 400, Al Ahsa 31982, Saudi Arabia \& Department of Mathematics and Statistics, College of Science, King Faisal University, P.O. Box 400, Al Ahsa 31982, Saudi Arabia. $^3$Email: sislam@kfu.edu.sa}}}

\affil[4]{{\scriptsize School of Physics, Damghan University, Damghan, 3671641167,  Iran. \& Center for Theoretical Physics, Khazar University, 41 Mehseti Street, Baku, AZ1096, Azerbaijan. $^{4}$Email: b.pourhassan@du.ac.ir}}

\affil[5,6]{{\scriptsize Department of Physics, College of Science,\\ King Faisal University, P.O. Box 400, Al Ahsa 31982, Saudi Arabia. $^{5}$Email: msadiq@kfu.edu.sa; $^{6}$Email: fharoon@kfu.edu.sa}}

\begin{document}

\date{}

\maketitle
\begin{abstract}
Anisotropic Bianchi type-III cosmology is examined within the framework of $f(R,T)$ gravity, 
where $R$ denotes the Ricci scalar and $T$ the trace of the energy-momentum tensor. In this work, we investigate the statistical constraints on anisotropic Bianchi type-III cosmology within the framework of $f(R,T)$ gravity. The specific choice $f(R,T)=R+2f(T)$ is considered and exact solutions are derived for the background dynamics of the model. The physical parameters, such as the Hubble parameter $H(z)$, spatial volume $V(z)$, energy density $\rho(z)$, and pressure $p(z)$, are derived and their evolutionary behaviors are analyzed. To examine the observational viability of the model, we employ Markov Chain Monte Carlo (MCMC) methods and perform a comprehensive statistical analysis using the latest observational datasets, including the Hubble parameter measurements, Baryon Acoustic Oscillations (BAO), and the Pantheon compilation of type Ia supernovae. The combined data analysis provides constraints on the free parameters of the model and allows a comparison with the standard $\Lambda$CDM cosmology. Our results show that the anisotropic Bianchi-III universe in $f(R,T)$ gravity can successfully accommodate current observational data, offering new insights into the role of matter–geometry coupling in the late-time cosmic acceleration.

\end{abstract}

\section{Introduction}
The analysis of cosmological observational data~\cite{1riess1998observational,2perlmutter2003measuring,3gold2009five,59dhankar2025observational} 
confirms that the present-day universe is undergoing an accelerated phase of expansion. This late-time acceleration was first established 
through high-redshift type Ia supernovae observations~\cite{1riess1998observational,2perlmutter2003measuring,4bennett2003microwave}. 
Additional evidence supporting this scenario arises from measurements of cosmic microwave background (CMB) radiation~\cite{5spergel2003first,6spergel2007three} 
and large-scale structure surveys~\cite{7tegmark2004cosmological}. Collectively, these observations indicate that the main driving component 
behind accelerated expansion is dark energy (DE)~\cite{48talole2023viscous,52khadekar2025holographic}, 
a mysterious form of energy characterized by negative pressure (or equivalently, positive energy density) that counteracts gravitational attraction. 
The second dominant component is dark matter, which plays a crucial role in the formation of cosmic structures. 
Current estimates suggest that the universe is composed of approximately $73\%$ dark energy, $23\%$ dark matter, and only $4\%$ baryonic matter. 
Although DE is widely accepted as the cause of cosmic acceleration, its true nature remains one of the greatest challenges in modern cosmology. 
Among several proposed candidates, the cosmological constant $\Lambda$ or vacuum energy stands out as the simplest explanation, 
being well supported by observations. Nonetheless, the $\Lambda$CDM model faces long-standing issues such as the fine-tuning and coincidence problems~\cite{8peebles2003cosmological,9astier2006supernova}.

Within the framework of general relativity (GR), the problem of cosmic acceleration can be addressed primarily via two approaches. 
The first involves modifying the matter sector by introducing various scalar fields, such as quintessence, phantom fields, tachyon fields~\cite{57samanta2025tachyonic}, 
and Chaplygin gas models~\cite{10sahni20045,11padmanabhan2008dark,12caldwell2002phantom,40khadekar2019modified,42georgiev2020two,49khadekar2019frw,54munyeshyaka2024perturbations,55thakran2025cosmological,58dhankar2025testing,61dhankar2025observational}.  
The second approach consists of modifying GR itself by altering the geometrical part of the Einstein--Hilbert action. 
Prominent alternatives include $f(R)$ gravity, $f(T)$ gravity~\cite{47gadbail2022interaction}, $f(R,G)$ gravity, $f(R;L)$ theory \cite{rudra},
and Gauss--Bonnet gravity~\cite{13nojiri2007introduction,14ferraro2007modified,15carroll2005cosmology,56munyeshyaka2025matter,60dhankar2025constraints,63dhankar2025testing}. Among these, $f(R)$ gravity has attracted particular attention due to its cosmological relevance. A further generalization, known as $f(R,T)$ gravity, was proposed by Harko et al.~\cite{16harko2011f}, where the Lagrangian depends on both the Ricci scalar $R$ 
and the trace $T$ of the energy--momentum tensor~\cite{62dhankar2025large}.  
Several works have explored particular functional forms of $f(R,T)$ within Friedmann-Robertson-Walker (FRW) cosmologies. 
For instance, Houndjo~\cite{17houndjo2012reconstruction} reconstructed $f(R,T)$ gravity by considering the function as a sum of two arbitrary functions of $R$ and $T$. 
Moreover, $f(R)$ models have been shown to unify early-time inflation with late-time cosmic acceleration~\cite{13nojiri2007introduction,18chiba2007solar,19multamaki2006spherically,20shamir2010some,53kumar2025joint,64sanyal2025cosmic}. 

Anisotropic cosmological models, such as those based on Bianchi types, are also of interest as they provide insights into the large-scale structure of the early universe. 
Yilmaz et al.~\cite{21yilmaz2012quark} investigated Bianchi-type spacetimes with quark and strange quark matter~\cite{46kumbhare2022strange} in the context of $f(R)$ gravity.  
Sharif and Zubair~\cite{22sharif2012anisotropic,23sharif2014study} analyzed perfect fluid and scalar field scenarios, studying energy conditions in Bianchi-type universes within $f(R,T)$ gravity.  
Other authors~\cite{24myrzakulov2012frw,25shaikh2017bianchi,26chaubey2013new,27mishra2014bianchi} have also investigated Bianchi-type models with perfect fluids
in the $f(R,T)$ framework~\cite{44islam2021gravitational,51panda2024thermodynamics}.  
Five-dimensional Kaluza--Klein cosmologies with perfect fluids have similarly been considered~\cite{28reddy2012kaluza,29ram2013some,50islam2024string}.  
In this context, Sahoo et al.~\cite{30sahoo2016kaluza} derived an effective cosmological constant $\Lambda$ depending on the stress-energy tensor $T$. 
Further studies include Bianchi type V cosmologies with $f(R,T) = f_1(R)+f_2(T)$~\cite{31ahmed2014bianchi}, bulk viscous models~\cite{32mahanta2014bulk}, 
and investigations of singularities in LRS Bianchi-type universes~\cite{33sahoo2015lrs,45kumar2022two}.

The accelerated expansion of the universe has posed significant challenges to the standard cosmological model, prompting the exploration of alternative theories beyond $\Lambda$CDM \cite{Riess1998, Perlmutter1999}. Among these, modified gravity theories such as $f(R,T)$-gravity, which extend General Relativity by incorporating both the Ricci scalar $R$ and the trace of the energy-momentum tensor $T$, have gained considerable attention as they offer potential explanations for dark energy and cosmic acceleration without the need for exotic matter fields \cite{16harko2011f, Nojiri2017}. Although the standard cosmological principle assumes spatial isotropy and homogeneity, observations of cosmic microwave background anisotropies and large-scale structures suggest the presence of small anisotropies that could influence cosmic dynamics \cite{Planck2018, Saadeh2016}. This has led to increased interest in anisotropic cosmological models, particularly those based on Bianchi classifications. The Bianchi-III model, characterized by anisotropic expansion and nonzero spatial curvature, provides a useful framework to investigate deviations from isotropy in the early and late-time universes \cite{Ellis1969, Collins1973}. Incorporating anisotropic Bianchi-III geometry within $f(R,T)$-gravity allows for a richer description of cosmic evolution, capturing both modified gravitational effects and anisotropic features. However, these models introduce additional parameters and complexity, making analytical solutions challenging and require advanced statistical techniques for parameter estimation \cite{Sharif2014, Moraes2017}.

Markov Chain Monte Carlo (MCMC) methods have become essential tools in cosmology for efficiently exploring high-dimensional parameter spaces and providing robust statistical constraints based on observational data sets \cite{Lewis2002, Trotta2008, Thakran:2025xwo, Dhankar:2025qmd, Dhankar:2025gid, Dhankar:2025fno}. Applying MCMC to anisotropic $f(R,T)$ cosmological models enables systematic constraint of model parameters, helping to assess their viability and consistency with current observations. Therefore, this study aims to employ MCMC methods to statistically constrain anisotropic Bianchi-III cosmologies in the framework of $f(R,T)$ gravity, with the goal of deepening our understanding of the interplay between anisotropy, modified gravity, and cosmic acceleration.

Motivated by these developments, we focus on spatially homogeneous Bianchi type III cosmological models in the framework of $f(R,T)$ gravity with perfect fluid.  
We consider a specific forms of the function: $f(R,T)=R+2f(T)$. The structure of the paper is as follows: In Section.~2, we derive the basic field equations of $f(R,T)$ gravity for a Bianchi type III metric with perfect fluid.  
Exact solutions of these field equations are presented in Section.~3. In this section, we explore the physical parameters of the model in more detail. The viability of our proposed model is examined using the latest observational Hubble parameter (OHD) measurements. In Section 4, Hubble, BAO and Pantheon data are used in model fitting. The physical implications of these solutions are discussed in detail in this section. Finally, in Section.~5 we summarizes the key results and conclusions of the work.

\section{Background of \texorpdfstring{$f(R,T)$}{f(R,T)}}
We have considered Bianchi Type-III metric in the form
\begin{equation}
    ds^2=dt^2-D^2dx^2-E^2e^{-2nx}dy^2-W^2dz^2,
\label{1}
\end{equation}
where $D, E, W$ are functions of $t$ only and $n$ is a constant.

The field equations of $f(R,T)$ gravity are derived from Hilbert-Einstein variational principle. The $f(R,T)$ action
is given as
\begin{equation}
    S=\int\sqrt{-g}\left(\frac{1}{16\pi G} f(R,T)+L_m\right) d^4x,
\label{2}
\end{equation}
where $f(R,T)$ is an arbitrary function of Ricci scalar $R$ and the trace $T$ of the energy-momentum tensor $T_{ab}$ of the
matter source. $L_m$ is the usual matter Lagrangian density. The stress energy tensor $T_{ab}$ of matter source is given by
\begin{equation}
    T_{ab}=-\frac{2}{\sqrt{-g}}\frac{\delta(\sqrt{-g}L_m)}{\delta g^{ab}}
\end{equation}
and its trace is $T = g^{ab}T_{ab}$.

Here, we have assumed that the matter Lagrangian $L_m$ depends only on $g_{ab}$ rather than its derivatives. Hence, we
obtain
\begin{equation}
    T_{ab}=g_{ab}L_m-\frac{\partial L_m}{\partial g^{ab}}.
\end{equation}
The $f(R,T)$ gravity field equations are obtained by varying the action $S$ in eq. \eqref{2} with respect to $g_{ab}$.
\begin{equation}
    f_R(R,T)R_{ab}-\frac{1}{2}f(R,T)g_{ab}+(g_{ab\Box - \nabla_a \nabla_b})f_R(R,T)=8\pi T_{ab}-f_T(R,T)T_{ab}-f_T(R,T)\Theta_{ab},
\label{5}
\end{equation}
where
\begin{equation}
\Theta_{ab}=-2T_{ab}+g_{ab}L_m- 2g^{lm}\frac{\partial^2L_m}{\partial g^{ab}\partial g^{lm}}.
\end{equation}
Here $f_R(R,T)=\frac{\partial f(R,T)}{\partial R}, f_T(R,T)=\frac{\partial f(R,T)}{\partial T}, \Box \equiv \nabla^a \nabla_b, $ where $\nabla_a$ is the covariant derivative.

Contracting eq. \eqref{5}, we get
\begin{equation}
    f_R(R,T)R+3\Box f_R(R,T)-2f(R,T)=(8\pi -f_T(R,T))T-f_T(R,T)\Theta,
\label{7}
\end{equation}
where, $\Theta=g^{ab}\Theta_{ab}.$
From eqs. \eqref{5} and \eqref{7}, the $f(R,T)$ gravity field equations takes the form

\begin{multline}
f_R(R,T)\left(R_{ab}-\frac{1}{3}Rg_{ab} \right)+\frac{1}{6}f(R,T)g_{ab}\\=(8\pi- f_T(R,T))\left(T_{ab}-\frac{1}{3}Tg_{ab} \right)\\-f_T(R,T)\left(\Theta_{ab}-\frac{1}{3}\Theta g_{ab} \right)+\nabla_a \nabla_b f_R(R,T).
\label{8}
\end{multline}
The standard stress energy tensor for matter Lagrangian is given by

\begin{equation}
    T_{ab}=(\rho+p)u_au_b-pg_{ab}
\label{9}
\end{equation}
where $u^a = (0, 0, 0, 1)$ is the four-velocity vector in co-moving coordinate system satisfying $u^a
u_a = 1$ and $u^a
\nabla_b u_a = 0$.
Here $\rho$ and $p$ denote the energy density and pressure of the fluid, respectively. The trace of the energy momentum
tensor for our model is $T = \rho - 3p$. Moreover, there is no unique definition for matter lagrangian. Here we assume it
as $ L_m = -p$ which yields
\begin{equation}
    \Theta_{ab}=-2T_{ab}-pg_{ab}.
\label{10}
\end{equation}
It is noted that the field equations of $f(R,T)$ theory of gravity depends on the physical nature of the matter field
through $\Theta_{ab}$ . Different cosmological models of $f(R,T)$ gravity are possible depending on the nature of the source of
matter. Harko et al. \cite{16harko2011f} constructed three types of models as follows:
\begin{equation}
    f(R,T)=\begin{cases}
        R+2f(T)\\
        f_1(R)+f_2(T)\\
        f_1(R)+f_2(R)f_3(T).
    \end{cases}
\end{equation}
For Bianchi type-III universe here we considered the first case, $f(R,T) = R+2f(T)$. With the help of eqs. \eqref{9} and \eqref{10} in the first model $f(R,T)=R+2f(T)$ For the particular choice $f(T) = \lambda T$,the field eqs. \eqref{8} gives
\begin{equation}
    R_{ab}-\frac{1}{2}g_{ab}R=(8\pi +2\lambda)T_{ab}+\lambda (\rho-p)g_{ab}.
\label{12}
\end{equation}

In the present work, we consider the units $G = c = 1$. The spatial volume $V$ for the metric in eq. \eqref{1} is
\begin{equation}
    V=DEW
\end{equation}
The scalar expansion $\theta$ and shear scalar $\sigma$ are defined as
\begin{equation}
    \theta=\mu^i_{;i}=\frac{\dot{D}}{D}+\frac{\dot{E}}{E}+\frac{\dot{W}}{W}
\end{equation}
and 
\begin{equation}
    \sigma^2  = \frac{1}{2}\sigma_{ab}\sigma^{ab}=\frac{1}{3}\left[\left(\frac{\dot{D}}{D} \right)^2+ \left(\frac{\dot{E}}{E} \right)^2+ \left(\frac{\dot{W}}{W} \right)^2-\frac{\dot{D}\dot{E}}{DE}-\frac{\dot{E}\dot{W}}{EW}-\frac{\dot{W}\dot{D}}{WD}\right].
\end{equation}
Hereafter an overhead dot denotes ordinary derivative with respect to $t$. The mean Hubble parameter $H$ is
\begin{equation}
    H=\frac{1}{3}\left(\frac{\dot{D}}{D}+\frac{\dot{E}}{E} +\frac{\dot{W}}{W}\right),
\end{equation}
where $H_1=\frac{\dot{D}}{D}, H_2=\frac{\dot{E}}{E}$ and $H_3=\frac{\dot{W}}{W}$ are the directional Hubble parameters in the spatial directions $x, y$ and $z$,
respectively.

The deceleration parameter $q$ is known to be a measure of cosmic acceleration. Since there is no unique definition
for directional Hubble parameter, we consider the deceleration parameter as \cite{34rahaman2005cosmological,35tripathy2010anisotropic,36saha2012dark}
\begin{equation}
    q=-\frac{\ddot{V}V}{\dot{V}^2}.
\label{18}
\end{equation}
The sign of $q$ determines the behaviour of the universe. The positive value of $q$ corresponds to a decelerating model while
its negative value indicates acceleration. Here, we choose to go for a negative deceleration parameter in accordance
with the recent observational data supporting an accelerating universe.

The mean anisotropy parameter $A_n$ is defined as
\begin{equation}
    A_n=\frac{1}{3}\sum_{l=1}^3\left(\frac{H_l-H}{H} \right)^2.
\label{19}
\end{equation}
In terms of the metric potentials, the Ricci scalar $R$ for the Bianchi III metric is expressed as
\begin{equation}
    R=2\left(\frac{\ddot{D}}{D} + \frac{\ddot{E}}{E} + \frac{\ddot{W}}{W} +\frac{\dot{D}\dot{E}}{DE}+\frac{\dot{E}\dot{W}}{EW}+\frac{\dot{W}\dot{D}}{WD} - \frac{n^2}{D^2}\right).
\label{20}
\end{equation}
\section{Exact solutions for the choice f(R, T) = R + 2f(T)
}
The field equation \eqref{12} for metric \eqref{1} takes the form
\begin{equation}
    \frac{\ddot{E}}{E} + \frac{\ddot{W}}{W}+\frac{\dot{E}\dot{W}}{EW}=(8\pi+3\lambda )p-\lambda\rho
\label{21}
\end{equation}
\begin{equation}
    \frac{\ddot{D}}{D}+\frac{\ddot{W}}{W}+\frac{\dot{D}\dot{W}}{DW}=(8\pi+3\lambda )p-\lambda\rho
\label{22}
\end{equation}
\begin{equation}
    \frac{\ddot{D}}{D}+\frac{\ddot{E}}{E}+\frac{\dot{D}\dot{E}}{DE}-\frac{n^2}{D^2}=(8\pi+3\lambda )\rho-\lambda p
\label{23}
\end{equation}
\begin{equation}
    \frac{\dot{D}\dot{E}}{DE}+\frac{\dot{E}\dot{W}}{EW}+\frac{\dot{D}\dot{W}}{DW}-\frac{n^2}{D^2}=-(8\pi+3\lambda )\rho+\lambda p
\label{24}
\end{equation}
\begin{equation}
    \frac{\dot{D}}{D}-\frac{\dot{E}}{E}=0.
\label{25}
\end{equation}
Equation \eqref{25} yields
\begin{equation}
    D=k_{1}E,
\label{26}
\end{equation}
where $k_{1}$ is the integration constant. Without loss of generality, we consider $k_{1} = 1$ which yields
\begin{equation}
    D=E
\end{equation}
Using eq. \eqref{26}, eqs. \eqref{21}-\eqref{24} reduces to
\begin{equation}
    \frac{\ddot{D}}{D}+\frac{\ddot{W}}{W}+\frac{\dot{D}\dot{W}}{DW}=(8\pi+3\lambda )p-\lambda\rho
\label{28}
\end{equation}
\begin{equation}
    2\frac{\ddot{D}}{D}+\left(\frac{\dot{D}}{D} \right)^2-\left(\frac{n}{D} \right)^2=(8\pi+3\lambda )p-\lambda\rho
\label{29}
\end{equation}
\begin{equation}
    \left(\frac{\dot{D}}{D} \right)^2+2\frac{\dot{D}\dot{W}}{DW}-\left(\frac{n}{D} \right)^2=-(8\pi+3\lambda )\rho+\lambda p.
\label{30}
\end{equation}
From eqs. \eqref{28} and \eqref{29}, we get
\begin{equation}
    \frac{\ddot{D}}{D}-\frac{\ddot{W}}{W}+\left(\frac{\dot{D}}{D} \right)^2-\frac{\dot{D}\dot{W}}{DW}-\left(\frac{n}{D} \right)^2=0
\label{31}
\end{equation}
Equation \eqref{31} contains two unknowns $D$ and $W$ and demands an additional assumed condition to be solved completely. In other words, if either of $D$ and $W$ is considered as a known function of $t$ from some plausible physical conditions of
the present universe, then we will be in a position to get a viable cosmological solution from eq. \eqref{31}. In general, to
handle such situation, the relation between the metric potentials is assumed to be $D = W^m$ corresponding to the fact
that the shear scalar $\sigma$ be proportional to the scalar expansion $\theta$ to get a physically realistic model\cite{37reddy2012bianchi,38mishra2015pressure}. Chandel
and Ram \cite{39chandel2013anisotropic}have used the generation technique to find the solutions of field equations of Bianchi type-III universe
in $f(R,T)$ gravity. In order to obtain a physically realistic model, we have considered the power law assumption as
\begin{equation}
    W=t^m
\label{32}
\end{equation}
where $m$ is a positive constant i.e. $m > 0$.The positive nature of $m$ is in accordance with the observational findings
which predicts an expanding universe.

After multiplying $D^2W$ in eq. \eqref{31}, we get
\begin{equation}
    \frac{d}{dt}(-D^2\dot{W}+DW\dot{D})=n^2W.
\label{33}
\end{equation}
Integration of eq. \eqref{33} yields
\begin{equation}
    -D^2\dot{W}+DW\dot{D}=n^2\left(\int Wdt+k_2 \right),
\label{34}
\end{equation}
where $k_2$ is a integrating constant. Equation \eqref{34} can be written as
\begin{equation}
    \frac{d}{dt}(D)^2-\frac{2\dot{W}}{W}D^2=F(t),
\label{35}
\end{equation}
where
\begin{equation}
    F(t)=\frac{2n^2}{W}\left(\int Wdt+k_2 \right).
\label{36}
\end{equation}
Equation \eqref{35} admits the solution
\begin{equation}
    D^2=W^2\left(\int \frac{F(t)}{W^2}dt +k_3\right),
\label{37}
\end{equation}
where $k_3$ is integrating constant. Using eq. \eqref{32} in eqs. \eqref{36} and \eqref{37}, we obtain
\begin{equation}
    D^2=\frac{n^2t^2}{1-m^2}+\frac{2k_2n^2t^{1-m}}{1-3m}+k_3t^{2m}.
\label{38}
\end{equation}
It is worth to mention, here, that the above solution for the directional scale factor $D$ is valid for the choices
$m\ne 1$. In principle, one can also obtain similar solution considering $m = 1$ or $W = t$ which may provide a logarithmic
universe. In the present work, we are more interested in the behaviour of the universe, where, the metric potential $W$
is considered as a function of cosmic time else than its linear behaviour and we have left the choice $m = 1$ or $W = t$ for
future investigation.

The deceleration parameter can now be obtained as
\begin{equation}
    q=-\frac{\left[\frac{n^2t^{2+m}}{1-m^2}+\frac{2k_2n^2t}{1-3m}+k_3t^{3m} \right]\left[\frac{(m+1)(m+2)n^2t^m}{1-m^2}+3m(3m-1)k_3t^{3m-2} \right]}{\left[\frac{(m+2)n^2t^{m+1}}{1-m^2} +\frac{2k_2n^2}{1-3m}+3mk_3t^{3m-1}\right]^2}.
\label{39}
\end{equation}
We are looking for a model explaining an expanding universe with acceleration, which for a suitable choice of $k_2$,
$k_3$ and $m$ gives the constant deceleration parameter. The current SNe Ia and CMBR observations favors accelerating
models $(q < 0)$. In view of this, we consider $k_2 = k_3 = 0$. From eqs. \eqref{38} and \eqref{39}, we obtain
\begin{equation}
    D^2=\frac{n^2t^2}{1-m^2}
\label{40}
\end{equation}
and 
\begin{equation}
    q= -\frac{m+1}{m+2}.
\label{41}
\end{equation}
It is obvious from \eqref{40} that, a physically acceptable scale factor can be obtained for $0 <m< 1$. In this range of the
parameter $m$, the deceleration parameter assumes a constant negative value as we have desired.
The metric \eqref{1} can be written as
\begin{equation}
    ds^2=dt^2-\frac{n^2t^2}{1-m^2}(dx^2+e^{-2nx}dy^2)-t^{2m}dz^2.
\label{42}
\end{equation}
This Bianchi type-III model represents an anisotropic cosmological universe in $f(R,T)$ gravity.

\subsection{Physical parameters of the model
}
The values of directional Hubble parameters $H_l (l = 1, 2, 3)$ are
\begin{equation}
    H_1=H_2=\frac{1}{t},  H_3=\frac{m}{t}.
\label{43}
\end{equation}

\begin{figure}
    \centering
    \includegraphics[width=0.75\linewidth]{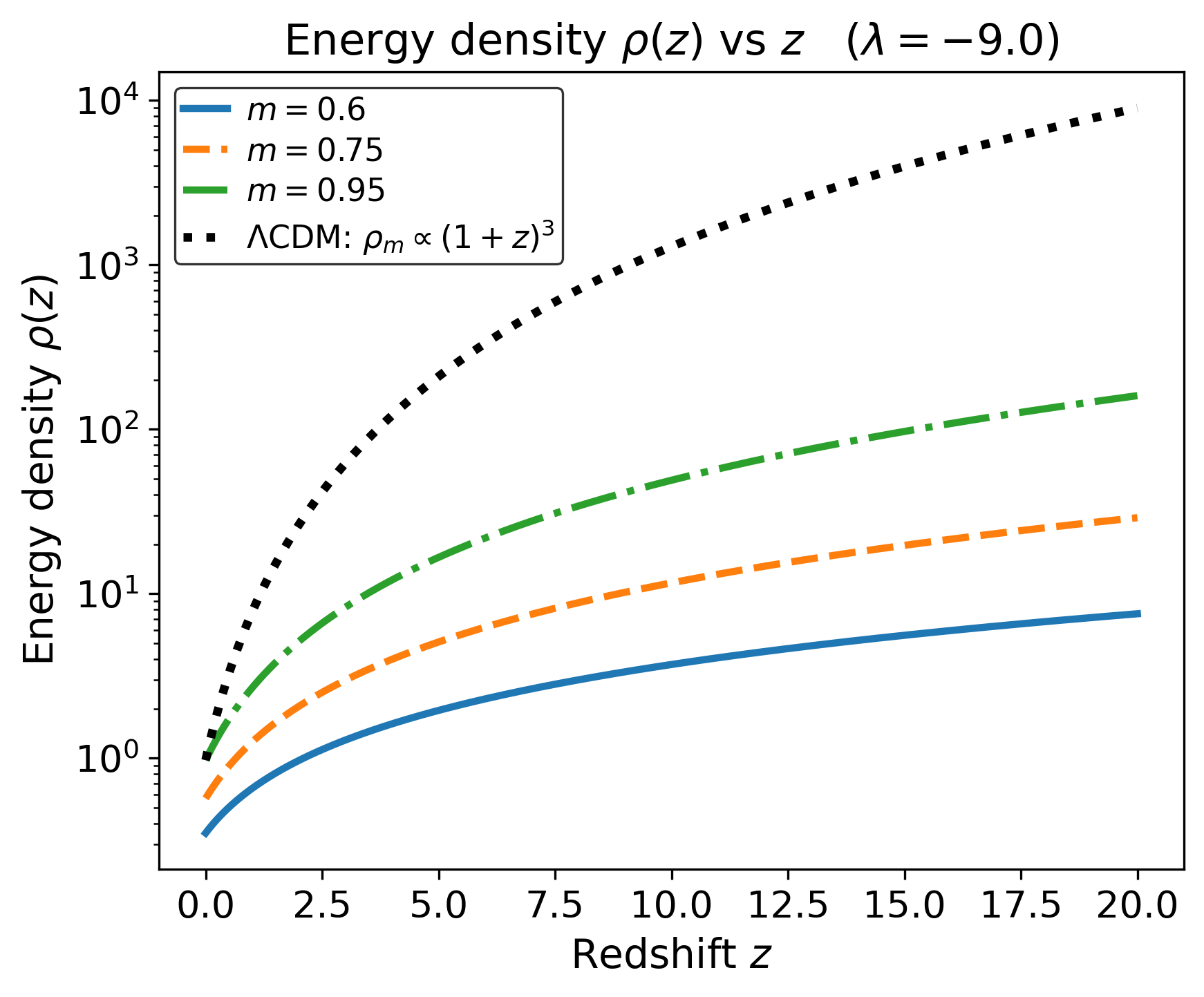}
    \caption{Energy density versus redshift ($z$) with $\lambda = -9$ and different values of m}
    \label{fig:1}
\end{figure}

\begin{figure}
    \centering
    \includegraphics[width=0.75\linewidth]{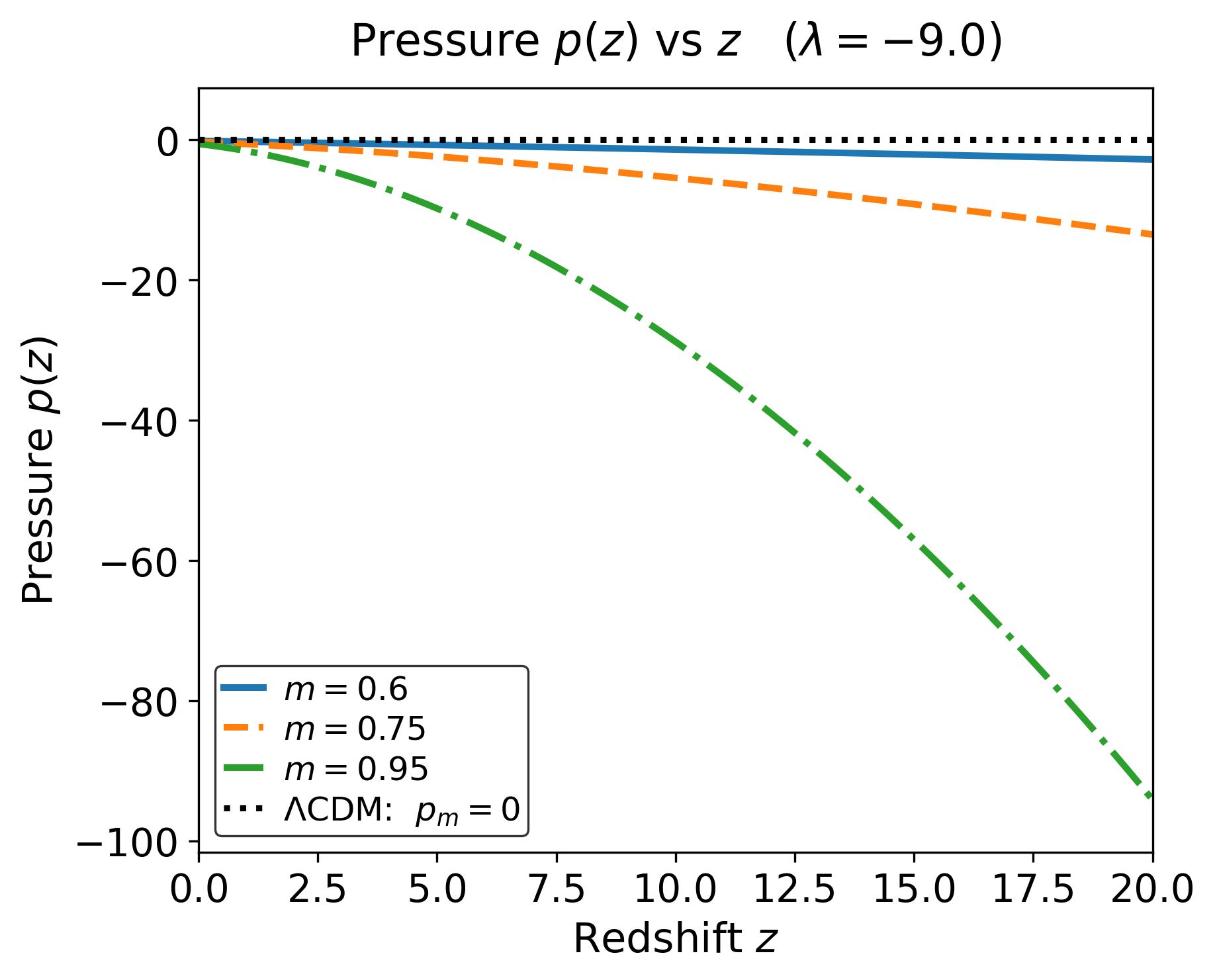}
    \caption{Pressure versus redshift ($z$) with $\lambda = -9$ and different values of m}
    \label{fig:2}
\end{figure}

\begin{figure}
    \centering
    \includegraphics[width=0.75\linewidth]{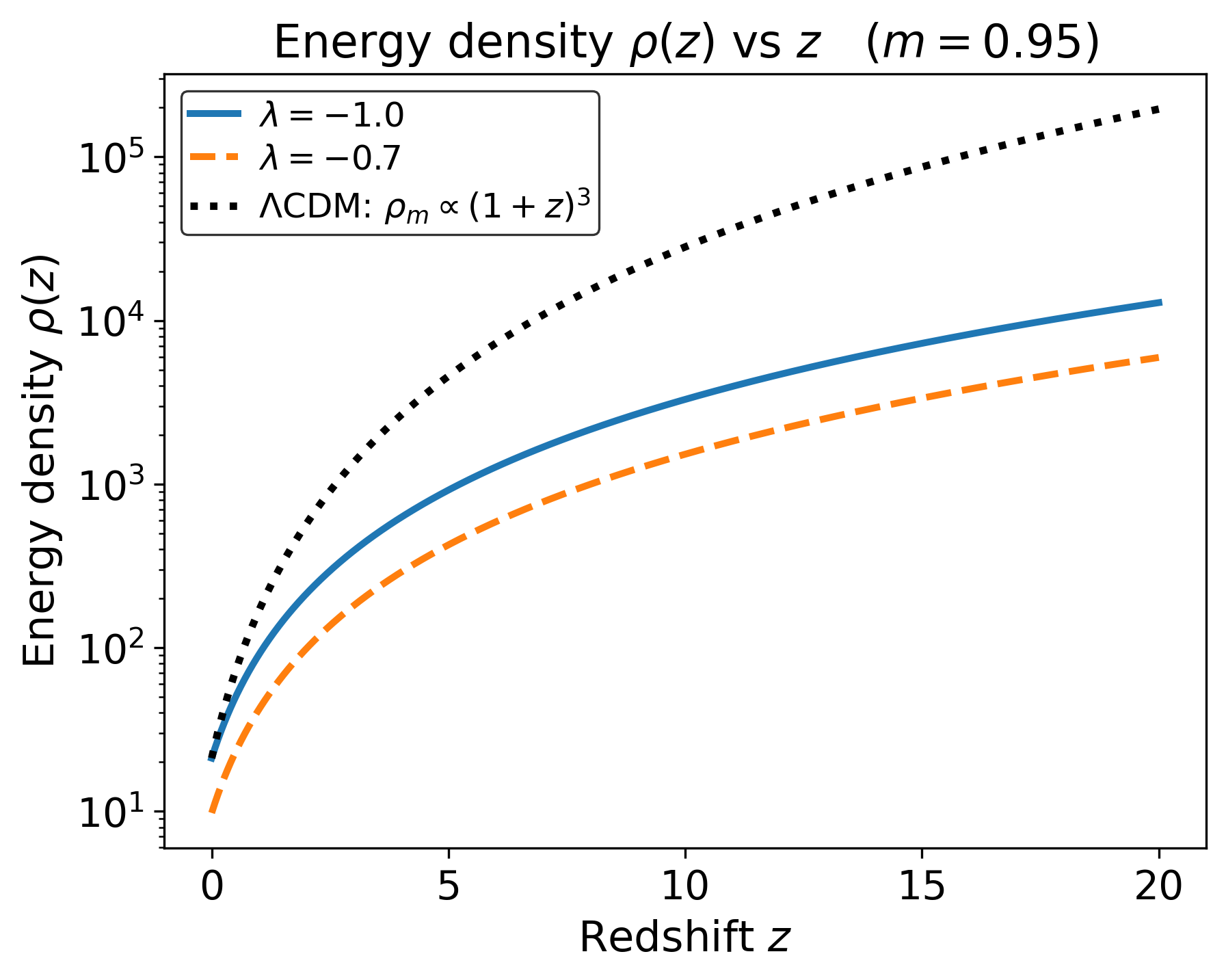}
    \caption{Energy density versus redshift ($z$) with $m=0.95$ and different values of $\lambda$ }
    \label{fig:3}
\end{figure}

\begin{figure}
    \centering
    \includegraphics[width=0.77\linewidth]{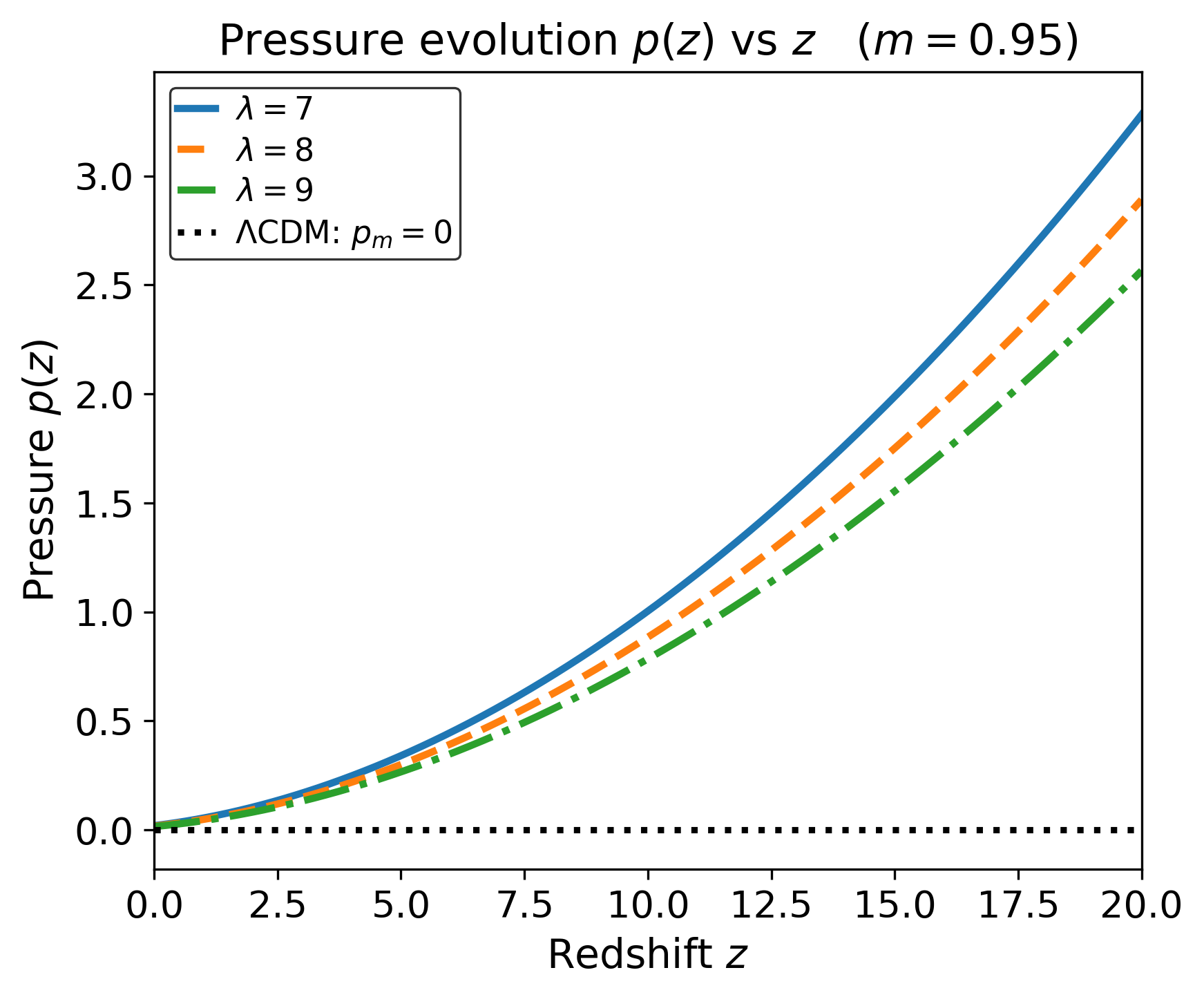}
    \caption{Pressure versus redshift ($z$) with $m=0.95$ and different values of $\lambda$}
    \label{fig:4}
\end{figure}

We have the $t-z$ relationship \cite{61dhankar2025observational}
\begin{equation}
    t(z)=\frac{1}{m\alpha}\log(1+(1+z)^{-m}),
\label{44}
\end{equation}
Using equation \eqref{44} in \eqref{43}, we get
\begin{align}
    H_1(z)=H_2(z)=\frac{1}{\frac{1}{m\alpha}\log(1+(1+z)^{-m})}=\frac{m\alpha}{\log(1+(1+z)^{-m})},
    \\ H_3(z)= \frac{m^{2}\alpha}{\log(1+(1+z)^{-m})}
\end{align}
\begin{equation}
    H=\frac{m+2}{3t}, V=\frac{n^2}{1-m^2}t^{m+2}.
\label{47}
\end{equation}
and using equation \eqref{44} in \eqref{47},we get the mean Hubble parameter $H$ as,
\begin{equation}
    H(z)=\frac{m+2}{3\frac{1}{m\alpha}\log(1+(1+z)^{-m})}=\frac{m\alpha(m+2)}{3\log(1+(1+z)^{-m})}
\end{equation}

\begin{equation}
    H(z) = H_0\left(\frac{\log2}{\log (1+(1+z)^{-m})}\right)
\end{equation}

and the volume $V$ is obtained as,
\begin{equation}
    V(z)=\frac{n^2}{1+m^2}\left(\frac{1}{m\alpha}\log(1+(1+z)^{-m})\right)^{m+2}
\end{equation}
From eq. \eqref{19} the anisotropy parameter $A_n $ becomes
\begin{equation}
    A_n=2\left(\frac{1-n}{2+n} \right)^2.
\end{equation}
The scalar expansion $\theta$ and the shear scalar $\sigma$ for the model \eqref{42} are given by
\begin{equation}
    \theta= \frac{m+2}{t},
    \sigma=\frac{1-m}{\sqrt{3}t}.
\end{equation}
The Scalar function $\theta$ and the shear scalar $\sigma$ in terms of $z$ are,
\begin{equation}
    \theta=\frac{m\alpha(n+2)}{\log(1+(1+z)^{-m})}, \sigma=\frac{m\alpha(1-m)}{\sqrt{3}\log(1+(1+z)^{-m})}
\end{equation}
The above equation shows the anisotropic nature of the Bianchi type-III model given in eq. \eqref{42}.
From eqs. \eqref{21}–\eqref{24}, the energy density of the model \eqref{42} is
\begin{equation}
    \rho=\frac{1}{t^2}\left[\frac{(8\pi+3\lambda)(2m+m^2)-\lambda m^2}{\lambda^2-(8\pi +3\lambda)^2} \right].
\end{equation}
\begin{equation}
    \rho(z)= \frac{m^2\alpha^2}{\left[\log(1+(1+z)^{-m})\right]^2}\left[\frac{(8\pi+3\lambda)(2\lambda+m^2)-\lambda m^2}{\lambda^2-(8\pi+3\lambda)^2}\right]
\end{equation}
and the pressure of the model becomes
\begin{equation}
    p= \frac{1}{t^2}\left[\frac{\lambda(2m+m^2)-(8\pi+3\lambda)m^2}{\lambda^2-(8\pi+3\lambda)^2} \right].
\end{equation}
\begin{equation}
    p(z)=\frac{m^2\alpha^2}{\left[\log(1+(1+z)^{-m})\right]^2}\left[\frac{\lambda(2m+m^2)-(8\pi+3\lambda)m^2}{\lambda^2-(8\pi+3\lambda)^2}\right]
\end{equation}
The plots of energy density $\rho$ and pressure $p$ against redshift coordinate $z$ for different values of $m$ and $\lambda$ are shown in
figs. (\ref{fig:1}),(\ref{fig:2}), (\ref{fig:3}) and (\ref{fig:4}), respectively. 
The equation of state (EoS) $\omega=\frac{p}{\rho}$ for the model \eqref{42} is obtained as
\begin{equation}
    \omega= \frac{\lambda(2m+m^2)-(8\pi +3\lambda)m^2}{(8\pi +3\lambda)(2m+m^2)-\lambda m^2}.
\end{equation}

\begin{figure}[H]
    \centering
    \includegraphics[width=0.75\linewidth]{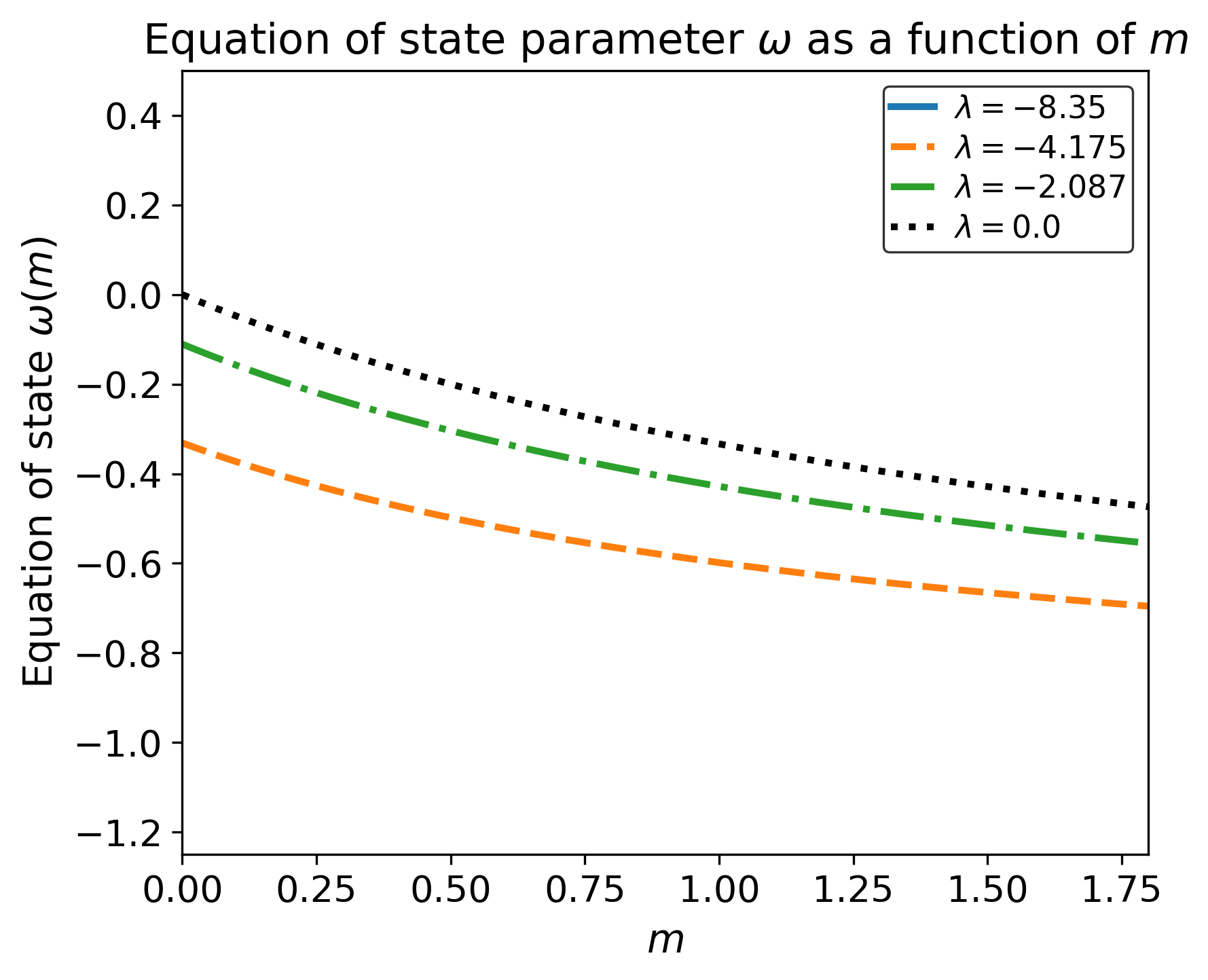}
    \caption{Equation of state parameter $\omega$ as a function of the model parameter m for some representative values of $\lambda$}
    \label{fig:5}
\end{figure}

It is interesting to note here that, for $\lambda = 0$, the $f(R,T)$ model reduces to general relativity (GR) and the EoS
$\omega$ becomes $\omega=-1+\frac{2}{m+2}.$ . Since for the present model, the model parameter m lies in the range $0 <m< 1$, the EoS
for general relativity should always be greater than $-\frac{1}{3}$ i.e., $\omega >-\frac{1}{3}$. The behaviour of the EoS for different choices
of $\lambda$ in the given range of $3$ is shown in fig. (\ref{fig:5}). It is evident, from the figure, that the choice of $\lambda$ has a great role in
deciding the behaviour of $\omega$. It is worth to mention here that the value of $\omega$ should lie in the negative range in accordance
with the recent observations predicting an accelerating universe and accordingly the parameter $\lambda$ can be constrained.
The Ricci scalar $R$ from eq. \eqref{20} of this universe is
\begin{equation}
    R= \frac{m(2m+1)}{t^2}
\end{equation}.
The scalar curvature R turns out to be zero for large time.

The trace $T$ of the model \eqref{42} is given by
\begin{equation}
    T=\frac{2}{t^2}\left[\frac{m(2m+1)(8\pi +3 \lambda)-\lambda m(2m+3)}{\lambda ^2-m^2} \right]
\end{equation}

\section{Observational Data Analysis}

In this work, the viability of the proposed model is examined using the latest observational Hubble parameter (OHD) measurements. 
\subsection{Hubble Data}
For this purpose, we employ a widely used compilation of 30 $H(z)$ data points obtained through the differential age (DA) technique, which provides estimates of the expansion rate of the universe at different redshifts. The Hubble function is defined as  
\begin{equation}
H(z) = -\frac{1}{1+z}\frac{dz}{dt},
\end{equation}
and is valid within the range $0.07 < z < 1.96$. To perform the statistical analysis, we minimize the chi-square function given by  
\begin{equation}
\chi^2_H = \sum_{i=1}^{30} \frac{\big[H_{\text{th}}(z_i) - H_{\text{obs}}(z_i)\big]^2}{\sigma^2_{H(z_i)}},
\end{equation}
where $H_{\text{th}}$ is the theoretical prediction, $H_{\text{obs}}$ the measured value, and $\sigma_{H(z_i)}$ the corresponding uncertainty for the $i$-th observational data point.

The considered Hubble dataset consisting of 30 data points are mentioned in the table below with the references.

\begin{table}[H]
    \centering
    \begin{tabular}{|c|c|c|c|l|l|l|l|l|l|l|l|}\hline
         z&  H(z)&  $\sigma_H$& Ref & z& H(z)& $\sigma_H$&Ref  & z& H(z)& $\sigma_H$&Ref  \\\hline
         0.070&  69&  19.6& \cite{zhang2014four} & 0.400& 95 & 17 &\cite{simon2005constraints} & 0.8754& 125& 17&\cite{moresco2012improved} 
\\\hline
         0.090&  69&  12& \cite{simon2005constraints} & 0.4004& 77 & 10.2 &\cite{moresco20166} &  
0.880& 90& 40&\cite{stern2010cosmic} 
\\\hline
         0.120&  68.6&  26.2& \cite{zhang2014four} & 0.4247& 87.1 & 11.2&\cite{moresco20166} & 0.900& 117& 23&\cite{simon2005constraints} 
\\\hline
         0.170&  83&  8& \cite{simon2005constraints} & 0.4497& 92.8& 12.9&\cite{moresco20166} &  
1.037& 154& 20&\cite{moresco2012improved} 
\\\hline
         0.1791&  75&  4& \cite{moresco2012improved} & 0.3802& 83 & 13.5  &\cite{moresco20166} & 1.300& 168& 17&\cite{simon2005constraints} 
\\\hline
         0.1993&  75&  5& \cite{moresco2012improved} & 0.4783& 80.9& 9&\cite{moresco20166}&  
1.363& 160& 33.6&\cite{moresco2015raising} 
\\\hline
         0.200&  7209&  29.6& \cite{zhang2014four} & 0.480& 97& 62&\cite{stern2010cosmic}& 1.430& 177& 18&\cite{simon2005constraints} 
\\\hline
         0.270&  77&  14& \cite{simon2005constraints} & 0.593& 104& 13&\cite{moresco2012improved}&  
1.530& 140& 14&\cite{simon2005constraints} 
\\\hline
         0.280&  88.8&  36.6& \cite{zhang2014four} & 0.6797& 92& 8&\cite{moresco2012improved}& 1.750& 202& 40&\cite{simon2005constraints} 
\\\hline
 0.3519&83 &14 & \cite{moresco2012improved} & 0.7812& 105& 12&\cite{moresco2012improved} 
&  
 1.965& 186.5& 50.4&\cite{moresco2015raising}  \\\hline
    \end{tabular}
    \caption{Hubble Parameter Values}
    \label{tab:1}
\end{table}

\begin{figure}[htbp]
    \centering
    \includegraphics[width=0.6\linewidth]{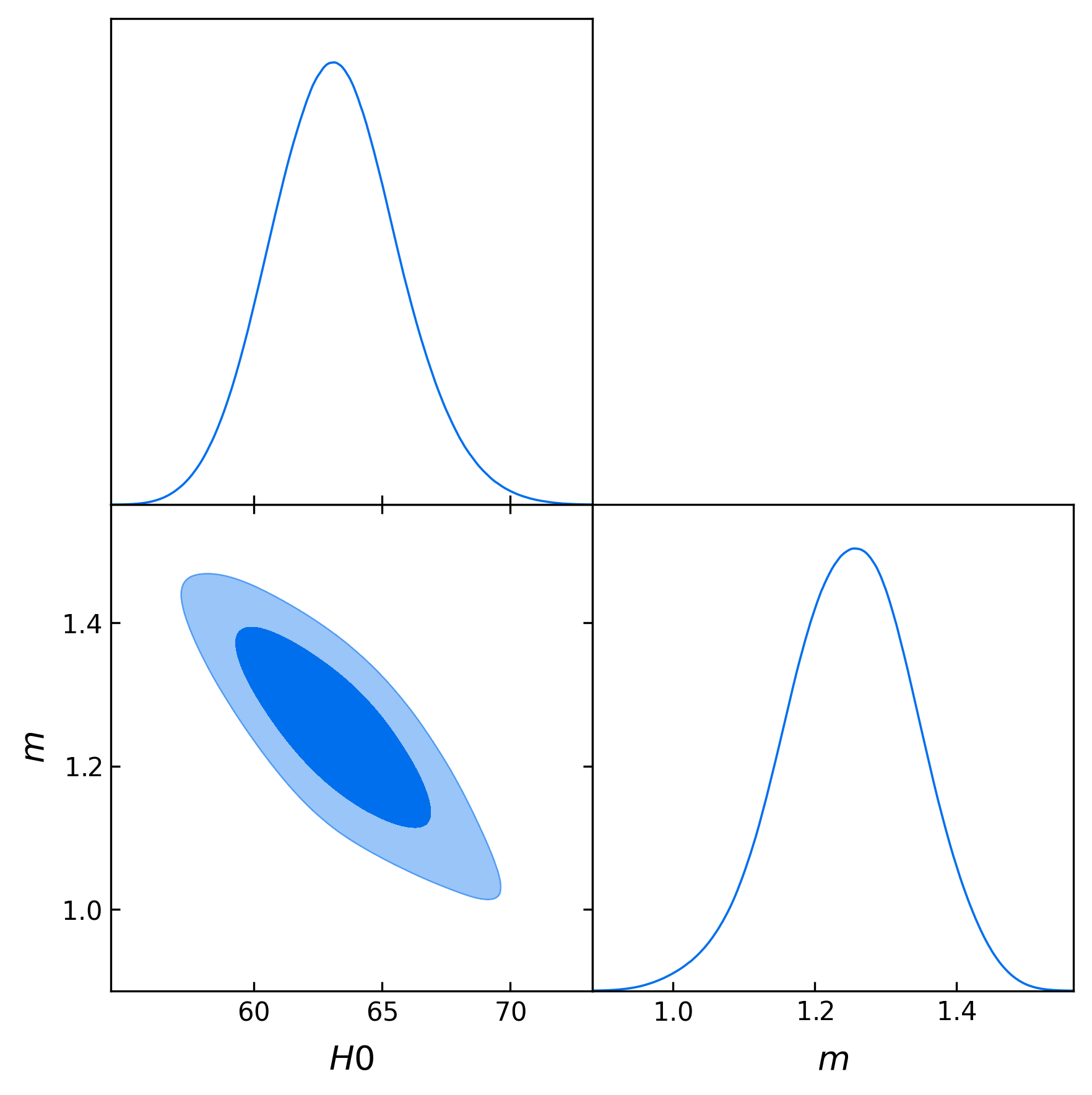}
    \caption{Model Fitting with Hubble Data}
    \label{fig:6}
\end{figure}

\begin{figure}[htbp]
    \centering
    \includegraphics[width=0.8\textwidth]{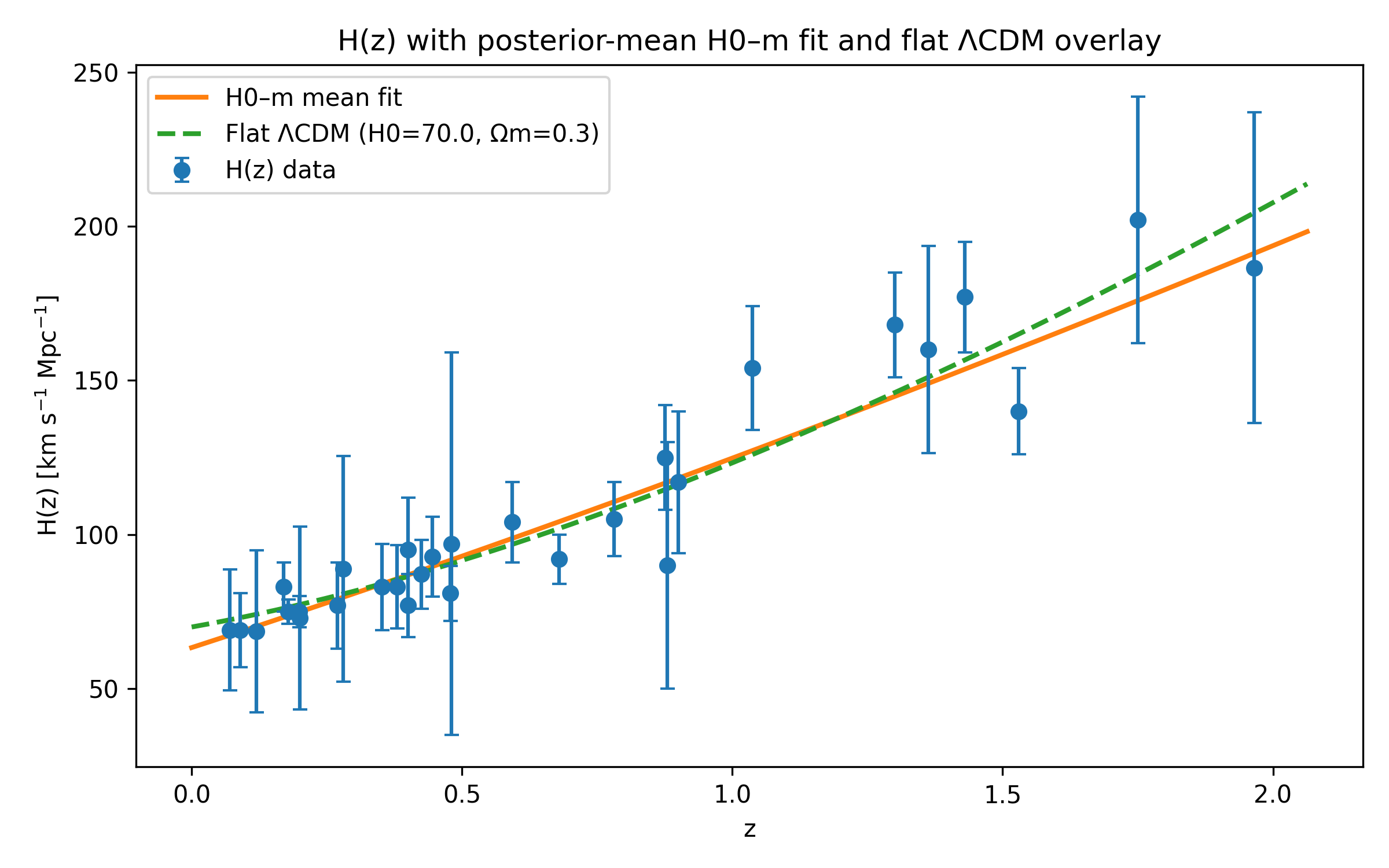}
    \caption{Error Plot with Hubble Data}
    \label{fig:7}
\end{figure}

\subsection{BAO Data}

We employ distance indicators constructed from the comoving sound horizon at the drag epoch, $r_d$, for Baryon Acoustic Oscillation (BAO) measurements. The range for the BAO data is $0.106 < z < 2.34$ . Based on the study and survey (isotropic versus anisotropic), the observed observable is among
\begin{equation}
\frac{D_V(z)}{r_d},\qquad 
\frac{D_A(z)}{r_d},\qquad 
\frac{D_H(z)}{r_d},\qquad 
\frac{r_d}{D_V(z)},\qquad 
\frac{r_d}{D_H(z)},\qquad 
\frac{H(z)\,r_d}{c}.
\label{eq:BAOobservables}
\end{equation}
The radial BAO distance scale is denoted by $D_H$, and the (transverse) angular diameter distance is denoted by $D_A$.
\begin{equation}
D_A(z)=\frac{D_C(z)}{1+z}, 
\qquad 
D_H(z)=\frac{c}{H(z)}, 
\qquad 
D_V(z)=\Big[z\,D_H(z)\,D_C^2(z)\Big]^{1/3},
\label{eq:DA_DH_DV}
\end{equation}
with the line-of-sight comoving distance
\begin{equation}
D_C(z)=c\int_{0}^{z}\frac{dz'}{H(z')}.
\label{eq:DC}
\end{equation}

For our phenomenological model we take
\begin{equation}
H(z;\boldsymbol\theta)=
\frac{m\,\alpha\,(m+2)}{3\ln\!\big(1+(1+z)^{-m}\big)}, 
\qquad \boldsymbol\theta=(m,\alpha,r_d),
\label{eq:H_malpha}
\end{equation}
and for the baseline flat $\Lambda$CDM comparison we use
\begin{equation}
H_{\Lambda{\rm CDM}}(z;H_0,\Omega_m)
=H_0\sqrt{\Omega_m(1+z)^3+1-\Omega_m}.
\label{eq:H_LCDM}
\end{equation}

Let $N_{\rm BAO}$ points with reported observable $X_i^{\rm (obs)}$ and covariance matrix $C_{ij}$ be present in the BAO catalogue at redshifts $\{z_i\}$.  Eqs.~\eqref{eq:DA_DH_DV}–\eqref{eq:H_malpha} are used to pick the appropriate expression from Eq.~\eqref{eq:BAOobservables}, which yields the corresponding theory prediction $X_i^{\rm (th)}(\boldsymbol\theta)$.  Next, the BAO chi-square is provided by
\begin{equation}
\chi^2_{\rm BAO}(\boldsymbol\theta)
=\big[X^{\rm (obs)}-X^{\rm (th)}(\boldsymbol\theta)\big]^T
\,C^{-1}\,
\big[X^{\rm (obs)}-X^{\rm (th)}(\boldsymbol\theta)\big].
\label{eq:chi2_BAO_cov}
\end{equation}
If the measurements are reported as uncorrelated with $1\sigma$ errors $\sigma_i$, we use the diagonal form
\begin{equation}
\chi^2_{\rm BAO}(\boldsymbol\theta)
=\sum_{i=1}^{N_{\rm BAO}}
\frac{\big[X_i^{\rm (obs)}-X_i^{\rm (th)}(\boldsymbol\theta)\big]^2}{\sigma_i^2}.
\label{eq:chi2_BAO_diag}
\end{equation}

When fitting the $(m,\alpha)$ model, the sound horizon $r_d$ is treated as a free nuisance parameter. For comparison, we also fit the $\Lambda$CDM baseline model using its own parameters $(H_0, \Omega_m, r_d)$ when BAO data are analyzed in isolation. Alternatively, if we wish to impose a consistent calibration for a fair comparison, we fit $(H_0, \Omega_m)$ while keeping $r_d$ shared between models. Throughout our analysis, we adopt $c = 299\,792.458~\mathrm{km\,s^{-1}}$, and all integrals are evaluated in units of $\mathrm{km\,s^{-1}\,Mpc^{-1}}$, resulting in distances expressed in megaparsecs (Mpc). It is worth noting that both isotropic and anisotropic constraints can appear among different BAO entries. In practice, we employ the covariance matrices provided by each survey—block-diagonal when the measurements are independent and combine the reported types according to Eq.~\eqref{eq:BAOobservables} to construct the final data vector. 

The BAO data points used for the data analysis is mentioned in the following table with the references.

\begin{table}[H]
    \centering
    \begin{tabular}{|c|c|c|c|c|l|l|l|l|l|}\hline
         Zeff&  Val&  error&  parameter& ref& Zeff& Val& error& parameter&ref\\\hline
         0.72&  2353& 63&  DVratio & \cite{bautista2018sdss} & 0.275& 0.1390& 0.0037& rdDV&\cite{percival2010baryon} 
\\\hline
         0.32&  1264&  25&  DVratio & \cite{tojeiro2014clustering} & 2.34& 8.86& 0.29 & DHrd&\cite{de2019baryon} 
\\\hline
         0.54&  9.212&  0.41&  DArd& \cite{seo2012acoustic} & 0.44& 0.0870& 0.0042& rdDV&\cite{blake2012wigglez}  
\\\hline
         0.106&  0.336 & 0.015  & rdDV  & \cite{beutler20116df} & 0.6& 0.0672& 0.0031& rdDV&\cite{nesseris2020evaporating} 
\\\hline
        0.15&  664  & 25.0 & DVratio & \cite{ross2015clustering} & 0.73
& 0.0593& 0.0020& rdDV&\cite{blake2012wigglez} 
\\\hline
         0.81&  10.75&  0.43&  DArd& \cite{abbott2019dark} &  1.480& 13.23& 0.47& DHrd&\cite{hou2021completed}\\\hline
         1.52&  3843&  147.0&  DVratio& \cite{ata2018clustering} & 0.697& 1499& 77& DAratio&\cite{sridhar2020clustering}\\\hline
         2.3&  34188&  1188&   Hxrd& \cite{delubac2013baryon}  & 0.874& 1680& 109& DAratio&\cite{sridhar2020clustering}\\\hline
         0.57&  13.67&   0.22&  DVrd& \cite{anderson2014clustering}  & & & & &\\\hline
    \end{tabular}
    \caption{BAO Datapoints}
    \label{tab:2}
\end{table}

\begin{figure}[htbp]
    \centering
    \includegraphics[width=0.6\textwidth]{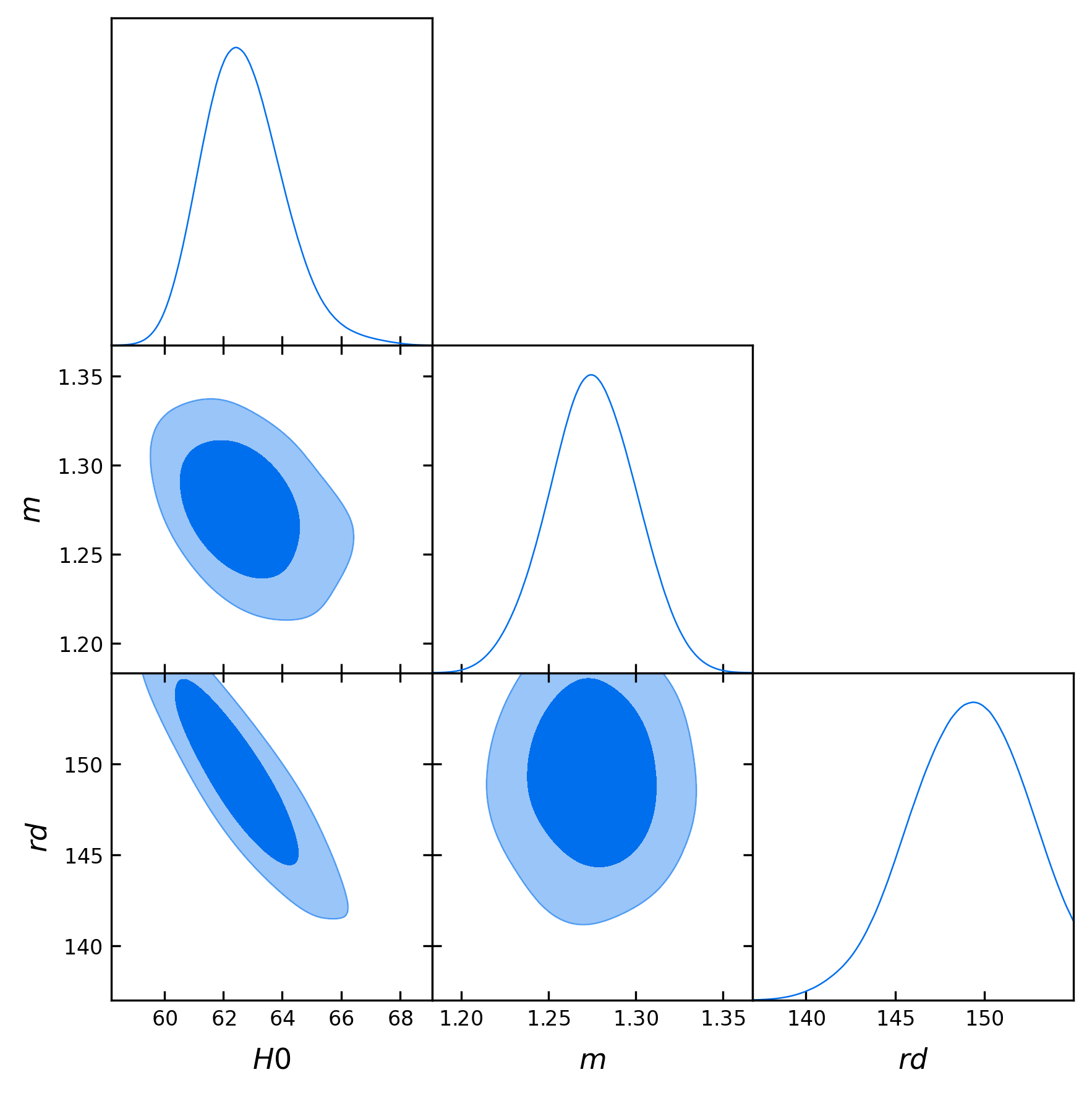}
    \caption{Model Fitting with Hubble and BAO Data}
    \label{fig:8}
\end{figure}

\begin{figure}[htbp]
    \centering
    \includegraphics[width=0.8\textwidth]{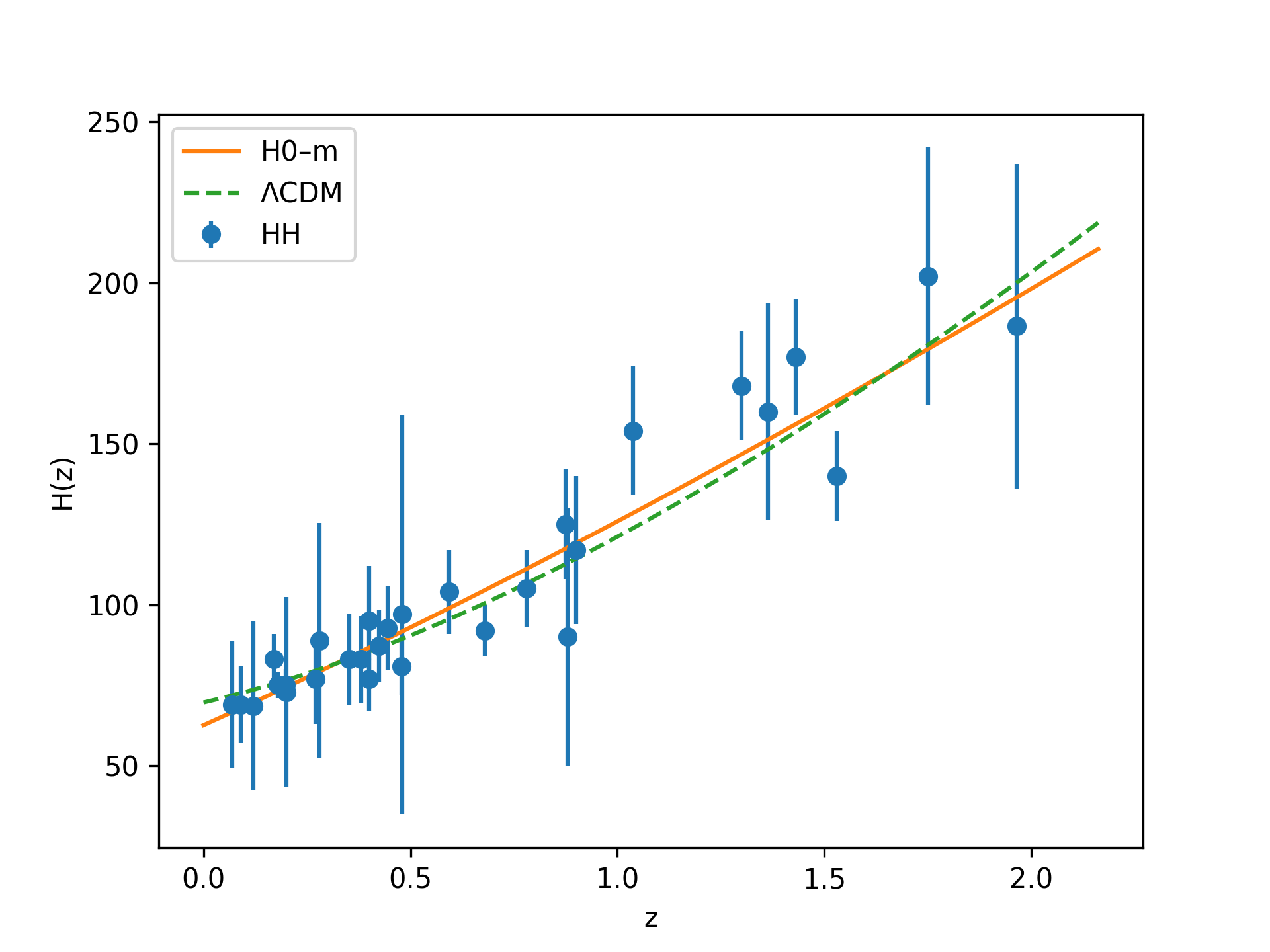}
    \caption{Error bar plot with Hubble and BAO data}
    \label{fig:9}
\end{figure}

\subsection{Pantheon data}
We employ 1048 datapoints from several sub-samples (PS1, SDSS, SNLS, low-$z$, and HST) covering the redshift range $0.01<z<2.26$ as standard candles from the Pantheon collection of Type Ia Supernovae (SNIa)\cite{scolnic2018complete}. The distance modulus is defined as\cite{chang2019constraining,zhao2019anisotropy}
\begin{equation}
\mu_{th}=5\log_{10}\frac{d_L}{Mpc}+25,
\end{equation} 
The Luminocity distance is given by $d_L= (c/H_0)D_L, H_0$ is the Hubble Constant, c is the speed of light and $D_L$ takes the form 
\begin{equation}
    D_L=(1+z_{cmb})\int_0^{z_{cmb}}\frac{dz}{E(z)}
\end{equation}
Where $z_{cmb}$ denotes the CMB frame redshift. The expression for $E(z)$ varies in different cosmological models. In the $\Lambda CDM $ model , $E(z)$ is given as 
\begin{equation}
    E^2(z)=\Omega_m (1+z)^3+(1-\Omega_m),
\end{equation}
Where $\Omega_m$ is matter density in the present epoch. The $\chi^2$ function for the Pantheon likelihood is defined as
\begin{equation}
\chi^2_{\rm SN}(\boldsymbol\theta) 
= \Delta \mu^T\,C_{\rm SN}^{-1}\,\Delta \mu,
\label{73}
\end{equation}
where $\Delta \mu = \mu^{\rm obs}-\mu^{\rm th}(\boldsymbol\theta)$, and $C_{\rm SN}$ is the covariance matrix of the dataset.  
The theoretical distance modulus is
\begin{equation}
\mu^{\rm th}(z;\boldsymbol\theta) = 5\log_{10}\!\left[\frac{D_L(z;\boldsymbol\theta)}{\mathrm{Mpc}}\right] + 25 + \mu_{\rm shift},
\label{eq:mu_theory}
\end{equation}
with $\mu_{\rm shift}$ included as a free nuisance parameter to account for the absolute magnitude/Hubble constant degeneracy.
The Pantheon dataset constrains only the form of $D_L(z)$ because of the degeneracy between $M$ and $H_0$; therefore, additional probes such as $H(z)$ or BAO are required for absolute calibration. When only diagonal errors are considered, Eq.~\eqref{73} reduces to a weighted sum with $\sigma_i^2$ appearing in the denominator. Whenever available, we make use of the full Pantheon covariance matrix, which incorporates both statistical and systematic uncertainties.

\begin{figure}[htbp]
    \centering
    \includegraphics[width=0.6\textwidth]{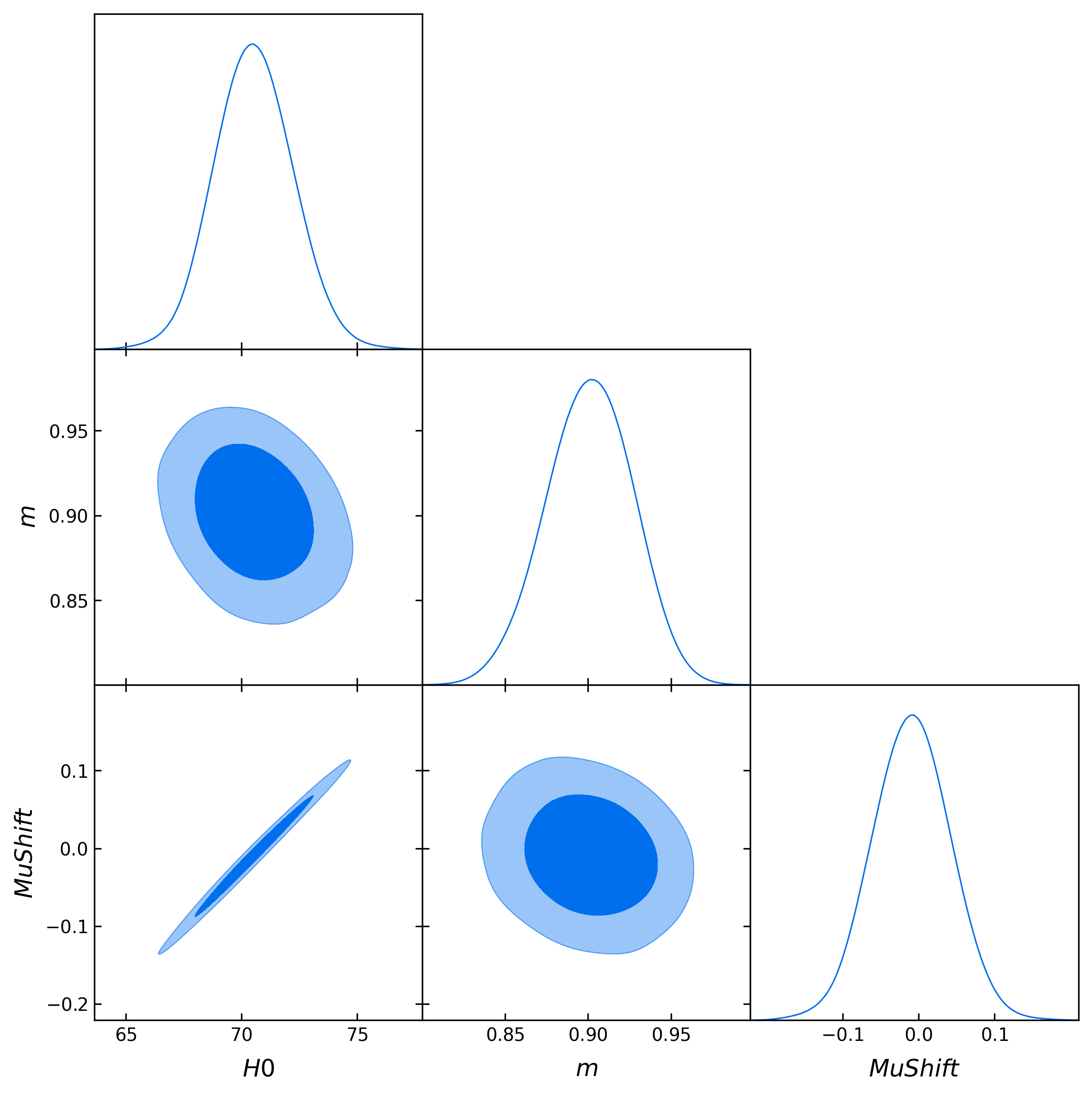}
    \caption{Model Fitting with Huble+Pantheon Data}
    \label{fig:10}
\end{figure}

\subsection{Model Free Parameters}
We adopt the priors as mentioned in table~(\ref{tab:3}), for our throughout data analysis process.

\begin{table}[H]
    \centering
    \begin{tabular}{|c|c|c|}\hline
         Free Parameter&  Symbol&  Prior Range\\\hline
         Hubble Constant&  $H_0$
&  $55\le H_0 \le 75$ Km/s/Mpc\\\hline
         Model Exponent&  $m$
&  $0.1\le m \le 2.0$\\\hline
         Sound Horizon Scale&  $r_d$
&  $135 \le r_d \le 155$ Mpc\\\hline
         Distance modulus offset&  $\mu_{shift}$&  $-1\le \mu_{shift} \le 1$\\ \hline
    \end{tabular}
    \caption{Model Free Parameters}
    \label{tab:3}
\end{table}

\begin{figure}[htbp]
    \centering
    \includegraphics[width=0.8\textwidth]{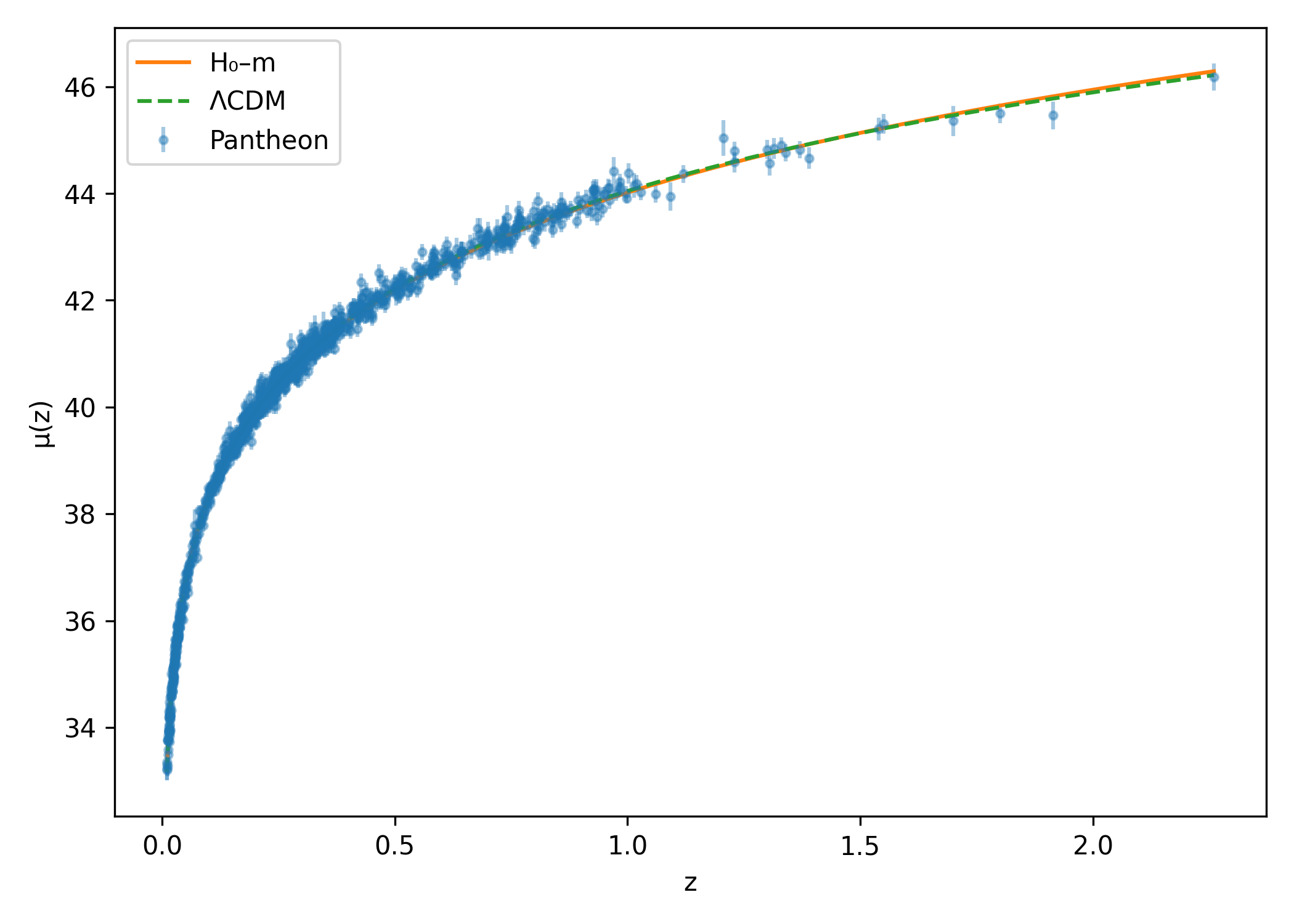}
    \caption{H(z) Error bar plot with Hubble and Pantheon Data}
    \label{fig:11}
\end{figure}

\begin{figure}
    \centering
    \includegraphics[width=0.6\textwidth]{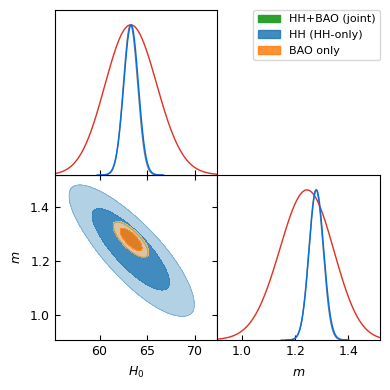}
    \caption{Model Fitting with Hubble, BAO And Hubble+BAO joint Analysis. }
    \label{fig:12}
\end{figure}

\begin{table}[H]
\centering
\begin{tabular}{|c|c|c|c|c|c|c|c|}
\hline
Data & Parameter & Bestfit Value & $\chi^2_{\text{red}}$& AIC & BIC & Del AIC & Del BIC \\
\hline
\multirow{2}{*}{H(z)} 
  & $H_0$     & 63 &  & & & &    \\ \cline{2-3}
  & $m$& 1.2
  & 0.5  & 19 & 22 & 4 & 7 \\ 
\hline
\multirow{3}{*}{H(z)+BAO} 
  & $H_0$     & 62.5 & & & & &   \\ \cline{2-3}
  & $m$& 1.2 & 0.6 & 35 & 40 & 5 & 5  \\ \cline{2-3}
  & $r_d$   & 148 &  & & & &  \\ 
\hline
\multirow{2}{*}{H(z)+Pth} 
  & $H_0$     & 70 &  & & & &    \\ \cline{2-3}
  & $m$& 0.9 & 1 & 1096 & 1111 & 39 & 39 \\ \cline{2-3}
  & $\mu_{shift}$ & -0.006 & & & & &\\
\hline
\end{tabular}
\caption{MCMC Results}
\label{tab:4}
\end{table}

\section{Conclusion}
In this work, we have investigated the accelerated expansion of the universe 
within the framework of $f(R,T)$ gravity. We considered a specific form of the 
function, namely $f(R,T) = R + 2f(T)$, and obtained exact solutions of the 
Bianchi type-III universe in the presence of a perfect fluid with the particular 
choice of constants $\lambda = \tfrac{1}{2}$ and $\mu = 1$. The relation between 
the metric potentials is taken as $D = W^m$, corresponding to the assumption that 
the shear scalar $\sigma$ is proportional to the scalar expansion $\theta$, in 
order to construct a physically realistic model. Furthermore, we adopted the 
power-law assumption $W = t^m$, where the accelerated Bianchi type-III universe 
is realized for $0 < m < 1$.  

The physical parameters of the model reveal that the spatial volume $V$ vanishes 
at $t=0$, implying that the universe originates with zero volume and expands with 
cosmic time $t$. The parameters $H$, $\theta$, and $\sigma$ diverge at the 
initial singularity but approach zero as $t \to \infty$, while the anisotropy 
of the universe persists throughout its evolution. The energy density $\rho$ 
and pressure $p$ diverge at the initial epoch and decay to zero at late times, 
ensuring the physical plausibility of the model.  

Thus, the models are physically realistic and viable for the following reasons:
\begin{itemize}
  \item As seen in Fig.~(\ref{fig:1}), the energy density $\rho(z)$ increases with redshift 
  $z$, taking higher values in the early universe and gradually decreasing as 
  the universe expands. For larger $m$, the growth of $\rho(z)$ is steeper, 
  with $m=0.95$ rising more rapidly compared to $m=0.6,0.75$, demonstrating the 
  sensitivity of the model to $m$ in the $f(R,T)$ framework with $\lambda=-9.0$. 
  In Fig.~(\ref{fig:2}), the pressure $p(z)$ decreases with redshift $z$, assuming more 
  negative values for higher $m$. This indicates an accelerated expansion phase, 
  where stronger negative pressure emerges as $m \to 1$, consistent with 
  dark-energy–like behavior and the dominance of repulsive gravity in the model 
  with $\lambda=-9.0$.

  \item Fig.~(\ref{fig:3}) shows that the energy density $\rho(z)$ increases with increasing 
  redshift $z$ for fixed $m=0.95$, and the incline becomes sharper for large negstive value of  
  $\lambda$. This reflects the sensitivity of $\rho(z)$ to the matter–geometry 
  coupling parameter $\lambda$, emphasizing its role in cosmic evolution and the 
  effective distribution of matter. Fig.~(\ref{fig:4}) shows that the pressure $p(z)$ grows 
  with redshift $z$, having higher values in the early universe and lower values 
  at late times. Increasing $\lambda$ reduces $p(z)$, highlighting the influence 
  of matter–geometry coupling on the anisotropic expansion. The equation of state parameter $\omega$ in Fig.~(\ref{fig:5}) decreases for all the values of coupling parameter $\lambda$ as $m$ increases. More negative values of $\lambda$ give a less value for $\omega$. This pattern demonstrates larger values of $m$ shifts the effective fluid deeper into the dark energy regime. This highlights the strong dependence of cosmic dynamics on the matter-geometry coupling parameter $\lambda$. 

  \item In Fig.~(\ref{fig:6}), the model fitting with Hubble data constrains the anisotropic 
  Bianchi-III cosmology in $f(R,T)$ gravity. The error bar plot with Hubble data 
  in Fig.~(\ref{fig:7}) demonstrates the compatibility of the theoretical model with 
  observations. Fig.~(\ref{fig:8}) presents the joint fitting with Hubble and BAO data, 
  which strengthens the statistical bounds on the model parameters. The error bar 
  plot with Hubble and BAO data in Fig.~(\ref{fig:9}) highlights the consistency of the 
  anisotropic dynamics with observations. In Fig.~(\ref{fig:10}), the model fitting with 
  Hubble+Pantheon data supports the cosmological viability of the framework, 
  while Fig.~(\ref{fig:11}) illustrates the error bar plot with Hubble and Pantheon data, 
  demonstrating robustness against supernova observations. Finally, in Fig.~(\ref{fig:12}), the 
  model fitting with Hubble, BAO, and their joint analysis demonstrates the 
  reliability of the cosmological predictions.
\end{itemize}

As a result, our analysis shows that the anisotropic Bianchi-III universe 
within $f(R,T)$ gravity is consistent with current observational data, thereby 
providing valuable insights into the role of matter–geometry coupling in driving 
the late-time accelerated expansion of the universe.

\section*{Data Availability statement}
 This research did not yield any new data.

\section*{Conflict of Interest}
Authors declare there is no conflict of interest.

\section*{Funding}
This work was supported by the Deanship of Scientific Research, Vice Presidency for Graduate Studies and Scientific Research, King Faisal University, Saudi Arabia (Funding No: KFU253533).

\section*{Acknowledgments}
~PKD would like to thank the Isaac Newton Institute for Mathematical Sciences, Cambridge, for support and hospitality during the programme Statistical mechanics, integrability and dispersive hydrodynamics where work on this paper was undertaken. This work was supported by EPSRC grant no EP/K032208/1. Also, PKD wishes to acknowledge that part of the numerical computation of this work was carried out on the computing cluster Pegasus of IUCAA, Pune, India and PKD gratefully acknowledges Inter-University Centre for Astronomy and Astrophysics (IUCAA), Pune, India for providing them a Visiting Associateship under which a part of this work was carried out. 


\bibliography{reference}
\bibliographystyle{unsrt}
\end{document}